# Development of a national thrust test facility for electric propulsion at Robinson Research Institute, New Zealand

**Emile Webster (corresponding author) and Ben Mallett**

Robinson Research Institute Victoria University of Wellington 69 Gracefield Road, Lower Hutt 5010, New Zealand

Emile.Webster@VUW.ac.nz and Ben.Mallett@VUW.ac.nz

## Abstract

The Robinson Research Institute is developing a type of electric propulsion for spacecraft called an applied-field magneto-plasma-dynamic (AF-MPD) thruster. The applied field module of the thruster features a cryocooler-cooled high-temperature-superconducting (HTS) magnet generating central fields exceeding 1 Tesla. This paper reports on the progress to date of the thruster test facilities established at the Institute to enable development of the AF-MPD and electric propulsion in general. The vacuum facilities consist primarily of a 1 m diameter, 3 m$^3$ cylindrical chamber with a high-vacuum pumping speed, measured to be ~4200 L·s$^{-1}$ [Ar]. The chamber can be reduced from atmospheric pressure to ~1·10$^{-5}$ hPa in about 2 h, enabling a rapid testing turn-around time, and maintains 5·10$^{-4}$ hPa during thruster operation at typical mass flow rates of 5 mg·s$^{-1}$ [Ar]. A pendulum thrust-stand has been purpose built to withstands static thruster loads of up to 30 kg and measure thrust forces up to ~200 mN. It is also designed to be inherently insensitive to the unique challenges of measuring AF-MPD with superconducting magnets; strong magnetic fields, cryocooler vibration, and the thermal gradients. Initial thrust-stand testing, whilst implementing both water-cooling and a cryocooler showed it to be capable of measuring forces to a precision of ~1 mN and with an absolute accuracy of ±2.3 mN (full scale). An extensive suite of software tools has also been developed to coordinate the large set of instruments needed to run the vacuum facilities, the thrust-stand, HTS magnet and cryocooler in addition to the many other auxiliary- and logging systems. Various upgrades to the facilities are in progress or planned.

## 1. Introduction

Thrusters for satellites and spacecraft powered by electricity, known as electric propulsion thrusters (EPT), are now used ubiquitously in the space industry [13,19]. Examples of these thrusters currently operating on spacecraft include; gridded ion thrusters [3], vacuum arc thrusters [17], pulsed plasma thrusters [45], and by far the most common of them, hall effect thrusters [18]. These thrusters can only operate in the vacuum of space, and work by using electrical power to first ionize their propellant, and then accelerate it with electromagnetic fields. The thrust produced can be expressed as;

$$T = \dot{m} \cdot v_{ex} = \dot{m} \cdot g_0 \cdot I_{sp}$$

In this expression, $\dot{m}$ is the mass flow rate of the propellant, and $v_{ex}$ is its exhaust velocity (averaged in the direction of the thrust). Traditional thrusters powered by chemical reactions typically achieve $v_{ex} < 3$ km·s$^{-1}$, whereas EPT typically operate at $v_{ex} > 15$ km·s$^{-1}$ [13]. With a higher $v_{ex}$, a lower mass of propellent is needed to generate a given amount of thrust. The specific impulse, $I_{sp} = v_{ex} \cdot g_0^{-1}$ where $g_0 = 9.81$ m·s$^{-2}$, with the units of seconds, is a key performance metric of EPT used in the literature to describe the efficiency of propellant consumption.

Another important performance metric of EPTs is efficiency. Efficiency, $\eta$, is the fraction of input electrical power, $P_{in}$, converted into the kinetic energy (per-unit-time) of the propellant in the direction of the thrust – also referred to as the kinetic power of the jet;

$$\eta = 0.5 \, \dot{m} \, v_{ex}^2 P_{in}^{-1} = 0.5 \, T^2 \, \dot{m}^{-1} \, P_{in}^{-1}$$

EPTs tend to generate thrust in the range 1−100 mN, orders of magnitude lower than many chemical rockets, which corresponds to mass flow rates typically measured in mg·s$^{-1}$. EPTs also have low thrust-to-weight ratio, typically less than 1:1000. This makes accurate measurement of the thrust from an EPT challenging [31], yet critically important in the evaluation of their performance.

In 2020 the New Zealand Government funded a consortium led by the Robinson Research Institute (RRI) to develop a type of EPT called applied-field magneto-plasma-dynamic (AF-MPD) thrusters. Substantial research on AF-MPD thrusters has been carried out since 1970, see [6] and references therein, which has identified promising performance characteristics of the thruster compared with other forms of EPT, such as high thrust density [37], high power operation and efficient operation with a wide range of propellants. AF-MPD thrusters use a magnetic field generated by its AF module to improve the thruster's performance. Traditionally copper electro-magnets are used for the AF module, for example, [16,24], however superconducting electromagnets are now a viable alternative thanks to the commercial availability of long lengths of high-quality second-generation superconducting tape [23]. Despite requiring cryogenic temperatures to operate, superconducting magnets are smaller, lighter, and able to achieve higher magnetic fields when compared with traditional copper-based magnets. Access to higher magnetic fields is enabling tests of performance scaling relations developed on lower field data [2,6] to higher magnetic fields, and have already shown promising performance improvements [39,46].

The research program led by RRI builds off the institute's long heritage in superconducting research and development [15,20,42], to develop superconducting AF-MPD thrusters [1,30] and demonstrate superconducting magnet technology in space through the Heki mission [32]. An integral part of the thruster development are test facilities that can accurately measure the thrust they will produce in space. Developing these at RRI as a national EPT thrust testing facility is seen as a key part of the research program. Two key components of the test facility are the vacuum chamber, used to simulate the environment of space (Section 1.3), and the force measurement system (Section 1.4), and both have demands placed on them dependent on the type of EPT being investigated.

## 1.3 Vacuum technology and requirements

As might be expected, the conditions generated inside a vacuum chamber on earth do not faithfully reproduce the conditions of space, specifically the micro-gravity, very low pressures, and undesirable chamber wall effects. Gravitational effects cannot easily be mitigated so must be tolerated and considered when assessing an EPTs performance.

Space-like pressures can be achieved with high pumping rates, but the actual operational pressure at the EPT will be highly contingent on the quantity of gas being fed to it and the pumping configuration. For example, to maintain a 'high-vacuum' (typically understood as $\sim 10^{-6}$ hPa) with a typical gas feed of about 1 mg·s$^{-1}$ [Xe], the pumping rate must be $\sim$20,000 L·s$^{-1}$ [Xe]. Typical laboratory sized turbomolecular pumps offer $\sim$500 L·s$^{-1}$. As such, determining an acceptable background pressure that will allow meaningful EPT performance measurements is of practical importance and this has been investigated from first principles [34] and experimentally [6]. Background pressure effects are expected to impact the plume expansion and re-ingestion of gas by the thruster.

Chamber wall effects include reflection of the EPT exhaust and the electrostatic potential of the wall, both of which affect the expansion of the plume and therefore EPT performance. The first effect can be assessed to some degree by comparative performance tests in vacuum chambers of varying size, and the second by changing the electrical potential of the chamber with respect to the plasma potential.

When a magnet is added to an EPT, as with an AF-MPD thruster, there will be additional magnet-wall interactions, where the magnetic field lines become distorted in the far field, nearer the chamber walls. Such interactions are true even for stainless steel chambers having a low magnetic permeability (often due to residual cold work or deviations in crystal structure). In principle, the expected magnitude of this effect would be minor, provided plasma detachment [9] (whereby the plasma no longer follows the magnetic field lines) occurs before the chamber wall permeability distorts the magnetic field.

As a consequence of the aforementioned constraints, most EPT test facilities use a combination of chamber length and symmetry to best control variables associated with the plume development and interactions, i.e., the chambers tend to be longer than they are wide and are cylindrical [8]. It is assumed the EPT will be thrusting in an axial direction and mounted at one end to allow maximum plume development before contacting the chamber. To minimize reflections, it has been shown that angular baffles and/or ion-absorbent materials (sputtering targets) ought to be used, like those described here, [26].

Typical pumping methods include turbomolecular- and/or cryogenic-pumping. Cryo-pumps scale well to the high pumping speeds ($\sim 10^4$ L·s$^{-1}$) demanded by EPT and the gases typically used in their operation (e.g. Ar, Kr, Xe), whilst turbomolecular pumps effectively pump light gases. 'Large-scale' dedicated EPT test facilities can have pump speeds of $\sim 10^5$ L·s$^{-1}$ [8,26,27,33,40,43], with 'laboratory scale' EPT test facilities offering pump speeds of $\sim 10^4$ L·s$^{-1}$ [14,46].

## 1.4 Thrust-stand design methodology

The inherent problem with all EPT is their thrust-to-weight ratios, which are often lower than 1:1000. Therefore, a constraining factor in the ground testing and development of EPTs is the process of removing (nulling out) intrinsic self-weight.

The important task of measuring thrust generated by low thrust-to-weight EPTs is achieved during ground testing using direct or indirect methods.

Indirect methods involve force measurements of the plasma plume ejected by the EPT. These measurements may be of the force imparted to a target by the plume that are registered by optical methods [11,41] or by a temperature-controlled strain-gauge/load-cell [5,39]. Additionally, the ion density and velocity profile of the plume can be interrogated and used to calculate the force generated by the thruster [12,38]. In the present context, indirect plume measurements have the advantage of being less susceptible to interference from vibrations, thermal effects and the magnetic fields associated with HTS-EPTs because the measurement device is mechanically decoupled and located some distance from the HTS-EPT. However, the indirect nature of the thrust measurement introduces systematic uncertainty and error in the measurement of the thrust. For example, the target itself alters the thruster plume resulting in back pressure and unwanted reflections. The target will form a solid angle, and as such only represents some small portion of the plume. Therefore, scanning the plume via an actuated force sensor, able to move in two or three dimensions, will be required to get the most accurate results. Even if such scanning is available the integrated results may not be representative of the true thrust if the EPT is prone to time variant behavior.

Direct methods involve mounting the EPT on a thrust stand and measuring the reaction force generated during operation of the EPT. Direct methods are considered more accurate and able to better represent the real performance of an EPT in space when compared to indirect methods. Thrust stands are sophisticated measurement devices as they must be capable of measuring the low thrust-to-weight ratio of EPTs against the perturbing influence of the electrical wiring and propellant gas/cooling-fluid connections [29]. Thrust stands of various designs have been in use for several decades and were pioneered in the mid 1960's when interest in interstellar flight highlighted the utility of EPT [25]. The present global capability to accurately measure the thrust generated by EPTs has been the result of decades of research and development undertaken in laboratories such as DLR [26,27], NASA [33], and others [7,14,35,36].

The basic operating principle of thrust stands is typically that of a pendulum, either of a simple [4,22], inverted [28] or horizontal-torsional type [10,44]. The relative merits of each variety of pendulum thrust balance for the measurement of EPTs has been discussed and summarized here [31]. However, this discussion does not consider the unique issues associated with measurement of an HTS AF-MPD thruster.

The use of an HTS magnet in an EPT poses additional challenges when making thrust measurements. These challenges arise from the strong magnetic fields produced by the magnets as an inherent aspect of their functionality, and from effects relating to the cryogenic subsystems necessary to support the operation of the superconductor. These subsystems introduce associated thermal management constraints as well as mechanical vibrations resulting from the electrical cryocoolers used to achieve the required operating temperatures of the HTS magnets (<70 K).

### 1.5 Target Requirements

This paper will now describe the development of a *national thruster test facility* at Robinson Research Institute, designed to service the requirements of EPTs, and importantly, AF-MPD thruster characterization. The target design parameters are based on existing literature and the superconducting AF-MPD under development:

- Dankanich 2013 [8] and Randolph 1993 [34] suggest that the vacuum chamber should be able to maintain a pressure at or below $5 \cdot 10^{-5}$ hPa during EPT testing.
- These facilities are targeting thrusters operating at powers <5 kW, and which could generate up to 200 mN of thrust [1,13].
- The large extra weight introduced when incorporating an HTS magnet and cryocooler could result in thruster weights of >20 kg.
- The thrust-stand ought to be able to reject at least 750 W of waste heat, generated by a typical cryocooler (~ 200 W) and by a kW-class AF-MPD.

This paper first details the various vacuum systems, critical sub-components, automation systems and the thrust-stand that, when combined, make up the test facilities. This is then followed by an in-depth investigation and discussion of the thrust-stand, including results from early commissioning and performance testing, and ends with a discussion of future improvements to enhance the facility's capabilities.

## 2. Experimental equipment

### 2.1 Vacuum systems

The test facility is built around, and in, a ~3 m$^3$ custom vacuum chamber made from 304 stainless steel and is named GERALDINE, figure 1. The cylindrical chamber measures 3.5 m in length and 1 m internal diameter. Pressures inside the chamber below $1 \cdot 10^{-2}$ hPa are measured with a Pfeiffer cold-cathode gauge, which has a quoted accuracy of 30%. A roughing vacuum, $\sim 7 \cdot 10^{-3}$ hPa, is supplied by an oil-free multi-stage roots pump (Pfeiffer ACG600), whilst a high-vacuum is supplied by a (nominal) 1400 L·s$^{-1}$ turbomolecular pump (Pfeiffer ATH1603M) and a cryopump (Trillium CP16), with nominal pumping speed of 4000 L·s$^{-1}$ [Ar]. Automated, pneumatically driven high-vacuum valves regulate the sequential operation of the pumps on the chamber. When empty, measurements indicate a leak/out-gas rate of $Q_{leak} \sim 2 \cdot 10^{-4}$ hPa·L·s$^{-1}$ from the chamber, and a base pressure of ~10$^{-8}$ hPa. Fully instrumented and with a thruster installed, typical outgas rates are $\sim 1 \cdot 10^{-2}$ hPa·L·s$^{-1}$ after a several hours, falling to ~3x10$^{-3}$ hPa·L·s$^{-1}$ after several days. The pumping system enables a 2-hour vacuum-vent-vacuum cycle, allowing for rapid experimentation. The current plan is to commission additional, custom-built Meissner-trap pumps to significantly increase the speed at which Ar (and heavier gases, such as Kr and Xe) can be pumped from the chamber during EPT testing.

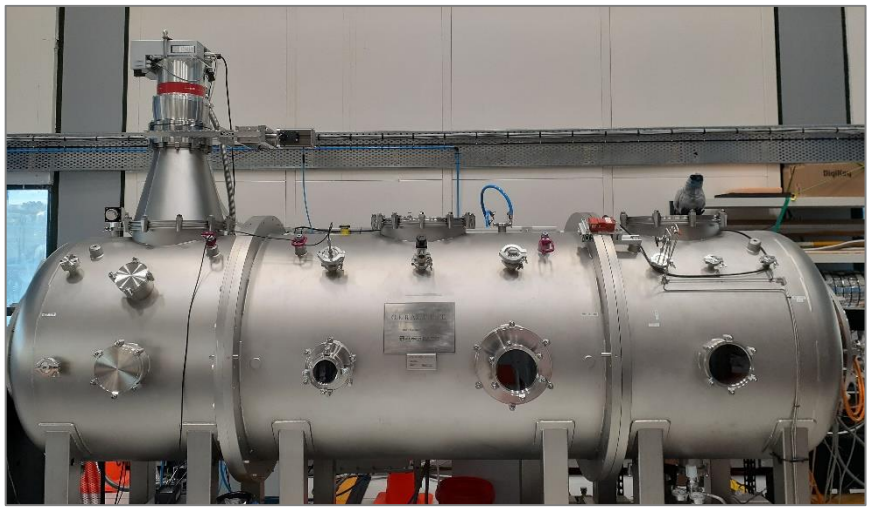

**Figure 1**, Three cubic meter vacuum chamber 'GERALDINE'.

Figure 2 shows data from an experiment to characterize the pumping speed on Ar of the ATH1603M turbomolecular pump, the CP16 cryopump and combined operation. In this experiment, Ar gas is admitted into the chamber by a calibrated mass-flow controller and the pressure monitored by the cold-cathode gauge. The pressure equilibrates after a few seconds after varying the mass-flow rate, $\dot{m}$, and that equilibrium pressure, P, is shown as a function of flow rate in the Figure. The pumping speed is S=Q/P, where the throughput is $Q=2.27 \cdot 10^4 \cdot \dot{m} \cdot M^{-1}$ with units [hPa·L·s$^{-1}$] where $\dot{m}$ is in [kg·s$^{-1}$] and the molar mass, M, of the gas species in [kg·Mol$^{-1}$]. These data show is the pumping speed is approximately constant over the range investigated and equal to 4200 +/- 100 L·s$^{-1}$ with both the turbomolecular and cryopumps running. As the temperature of the cold-stage of the cryopump is typically 10 K where the equilibrium vapor pressure of Ar is vanishingly low, it is expected that the pumping speed will be similar for the other noble gases Kr and Xe commonly used in electric propulsion.

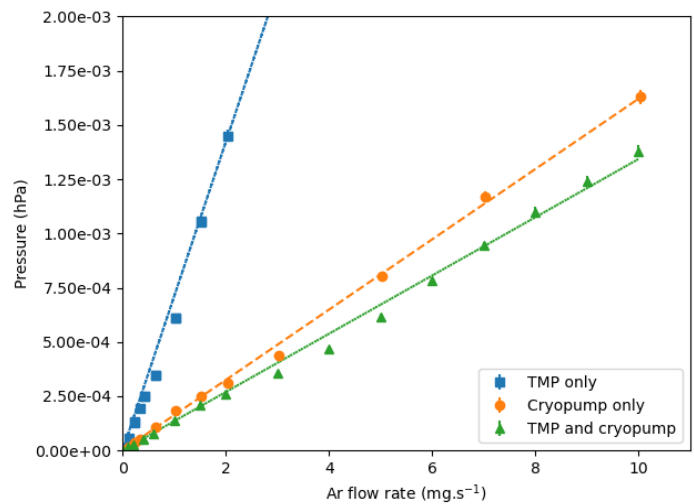

**Figure 2.** Measured equilibrium vacuum pressure as Ar gas is injected into the chamber under three pumping conditions: (i) with the ATH1603M turbo-molecular pump as the sole acting pump (blue squares), (ii) with the CP16 cryopump as the sole acting pump (orange circles), (iii) with both high-vacuum pumps (green triangles). The lines are best linear fits to the data, from which we determine the Ar pumping speeds of 800 L·s$^{-1}$, 3500 L·s$^{-1}$ and 4200 L·s$^{-1}$ respectively with an uncertainty of 100 L·s$^{-1}$.

## 2.2 Pendulum thrust-stand.
### 2.2.1 Pendulum design

Early in the development of the test facilities it was clear that some of the more elaborate and exotic thrust-balance configurations, for example, closed-loop-controlled double inverted pendulums [28] and complex interferometry setups [4,7] were not going to be feasible, given time and resource constraints. Even so, preliminary calculations showed that a carefully designed simple hanging pendulum could offer the required sensitivity, using either a loaded beam (static measurement) or a high-accuracy optical distancing device (dynamic measurement), both of which were implemented and tested. A hanging pendulum is one of the simplest thrust-stands to build and has the advantage of being intrinsically stable, with gravity providing the restorative force.

To maximize the sensitivity of both static and dynamic measurements the longest practicable swing arm length of ~0.5 m was chosen, which was determined by the internal diameter (~1 m) of the vacuum chamber. The main frame of the thrust stand is constructed from a standard 40 mm square engineering aluminum extrusion a with a 450 x 600 mm optical breadboard forming the base. The pendulum A-frame arms are made from 20 mm think aluminum, and the pendulum plate is a smaller optical breadboard measuring 300 x 450 mm. Custom machined aluminum bearing supports house the two main full-ceramic pivot bearings. Where possible all parts have been made of aluminum or brass to minimize magnetic coupling of critical pendulum components with those of the superconducting magnet. The initial computer aided design (CAD) of the thrust stand is shown in figure 3a along with two potential placements for a water chiller plate (Appendix, 6.2), one mounted on the underside of the pendulum plate and the other to the side in the vertical orientation (affixed to the rear of an additional 450 x 300 mm breadboard for mounting a cryocooler). The thrust-stand (as-built) is shown in Figure 3b sitting atop a telescopic rail system with pocketed soft-silicone vibration isolators.

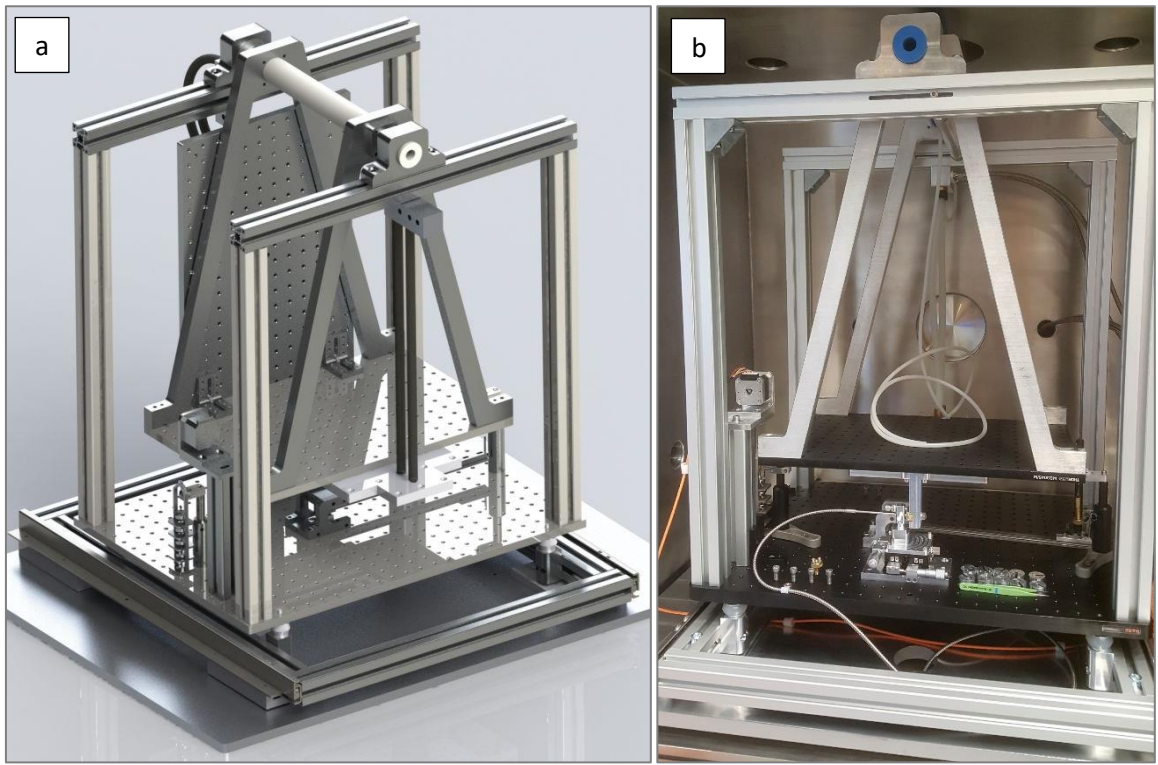

**Figure 3a**, CAD design of thrust-stand and, **b**, the as-built thrust-stand within a smaller 570 L development vacuum chamber ready for performance testing.

For a dynamic displacement measurement, the sensitivity falls as a function of the loading mass ($m$) and the pivot bearing static friction ($\mu$). A first order assessment of the frictional component can be made by combining the various torques in combination with calibration displacement data. Under equilibrium conditions, either constant thrust or weight calibration, the restoring torque ($\tau_{Restore}$) resulting from gravity ($g$) on the pendulum plate must be equal to the calibration torque minus the static friction,

$$R_{COM} \cdot m \cdot g \cdot \sin\theta = R_{cal} \cdot F_{cal} - R_{Bear} \cdot \mu \cdot m \cdot g,$$

where $R_{COM}$ is the center-of-mass radius, $R_{cal}$ is the weight calibration radius, $R_{Bear}$ is the bearing radius and $\mu$ is the coefficient of static friction for the two pivot bearings. The measured displacement ($d$) will be similarly defined as $d = R_{opt} \cdot \sin\theta$, where $R_{opt}$ is the optical displacement sensor radius. Hence, combining these two equations the displacement equation describing both the thrust force and fictional component can now be defined as,

$$d = R_{opt} \cdot R_{COM}^{-1} \cdot [F_{cal} \cdot R_{cal} \cdot (m \cdot g)^{-1} - R_{Bear} \cdot \mu],$$

Therefore, given known quantities of $d$, $F_{cal}$, and $m$, a value for the static friction can then be derived,

$$\mu = [R_{cal} \cdot F_{cal} \cdot (m \cdot g)^{-1} - d \cdot R_{COM} \cdot R_{opt}^{-1}] \cdot R_{Bear}^{-1},$$

which is useful for determining what proportion of potential thrust force measurement is lost to the bearing friction. The actual calculations based on key measurements are presented in Section 4.1.2 as part of the results and discussion.

For static measurements of force using a load-cell the relationship must consider not just the various force components that couple the pendulum plate and load-cell, but also a load-beam that provides mechanical amplification to the loadcell, as shown in figure 4. Hence, the sensitivity is still proportional to the loading mass, pendulum length, the pivot bearing diameter and its static friction contribution. However, the torque on the pendulum plate is now transferred through a small idler pulley via a carbon fiber thread to a 293 mm long I-profile aluminum load-beam. This load-beam is actuated about a hardened steel knife-edge pivot 13 mm from one end and provides an approximate twenty-fold increase in the force applied to the load-cell. The sensitivity of the dynamic and optical measurement systems for varying thruster masses are expected to be similar, given that the displacement angle of the load-beam must be small and that it is not subjected to the thruster mass atop the pendulum plate. Therefore, variations between the two systems should be solely attributable to frictional losses resulting from the extra idler pulley and knife-edge.

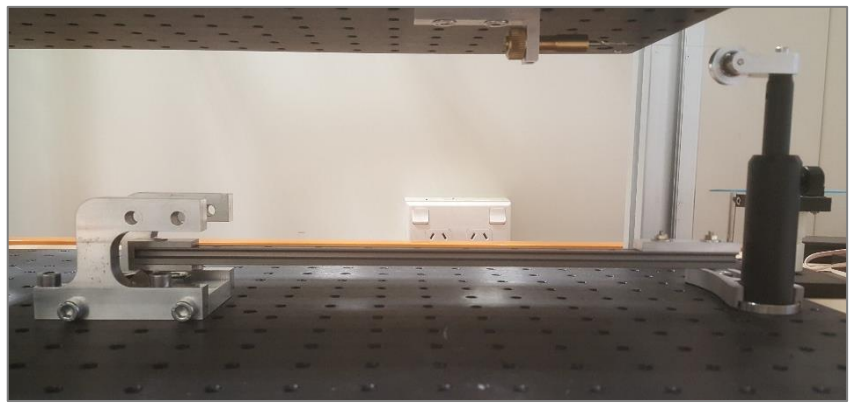

**Figure 4**, load-beam and pulley arrangement mounted below the pendulum plate, used to provide force amplification to the load-cell.

A simplified diagram, figure 5, shows how critical measurement systems are arranged on the thrust-stand. Also, shown is the thruster atop the pendulum plate and the arrangement of the load-beam. The radius distance between pivot and calibration bearing ($R_{calibration}$) was deliberately chosen to be the same as that between the pivot and the load-beam bearing. The thruster center-of-mass (COM), ideally, ought to coincide with the pendulum plate COM to maximize the thrust-stand's sensitivity and remove the need for a counterweight. One way of meeting this condition is to mount the thruster on an adjustable linear slide. In figure 5 the thruster is shown above the pendulum plate to allow for the various mounting and adjustment components that sit beneath it. This offset will lower the thrust-stand's sensitivity by the ratio of $R_{Thrust}/R_{calibration}$. Therefore, it is desirable for the thruster to be located as close to the pendulum plate as practicable.

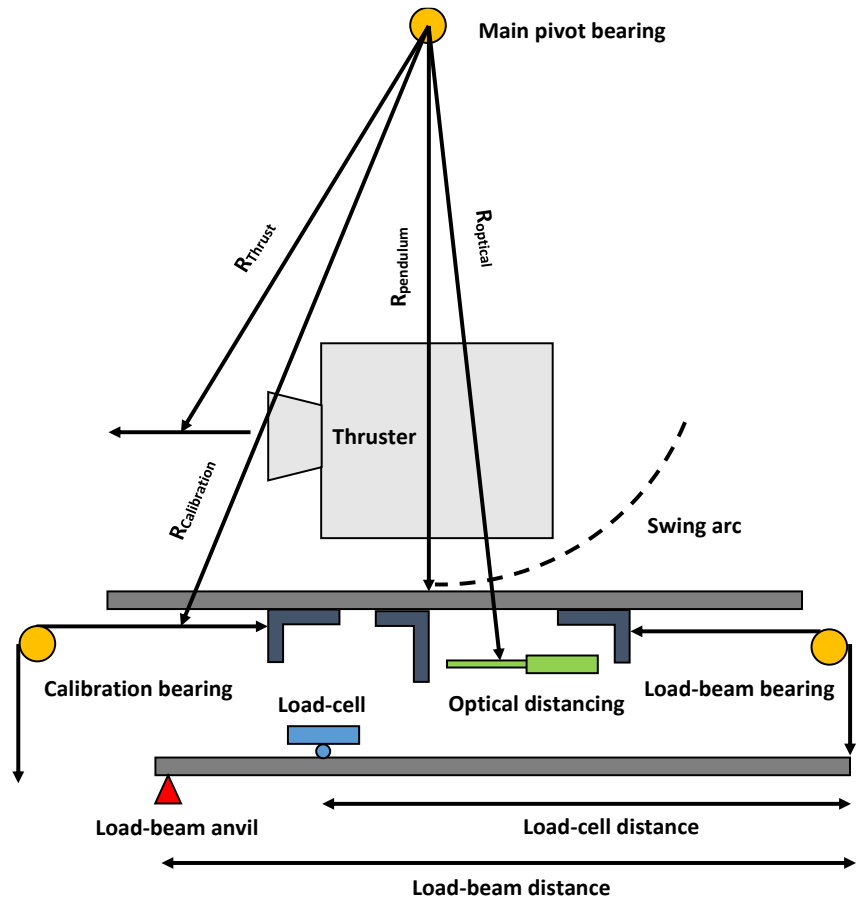

**Figure 5**, a simplified mechanical diagram of thrust-stand showing important radii.

A list of key distances that determine the thrust-stands sensitivity are provided in Table 1.

**Table 1**, of key distances, for determining forces and torques.

| Parameter | Value |
|---|---|
| $R_{calibration}$ | 554 mm |
| $R_{optical}$ | 592 mm |
| $R_{pendulum}$ | 525 mm |
| Load-cell distance | 280 mm |
| Load-beam distance | 293 mm |

### 2.2.2 Parasitic forces

To make the best use of the sensitivity available, for either dynamic or static thrust measurements, an important consideration was to limit the parasitic force loads that would further reduce sensitivity and add undesirable hysteresis effects (resulting in larger uncertainties). The design incorporates a pair of large diameter full-ceramic bearings (BOCA, 6006 SI3N4, 55 mm OD and 30 mm ID), shown to the left and right in figure 6a. These were ultrasonically cleaned to remove any residual oil (as is required for in-vacuum use) and to minimize the static and dynamic rolling friction. It was found that that an adjustable spreader-bar (aluminum bar sitting below pivot axel, Fig. 6a and 6b) was needed between the two main horizontal supports to ensure the two bearings remained precisely aligned (discussed further in Results Section 4.1.1), thereby minimizing bearing side-load friction. Part of the function of the unoccupied internal bearing area is to carry the electrical cabling and connections for: powering the thruster (3 x 20 A), the HTS magnet (2 x 30 A), the cryocooler (2 x 10 A), and up to 24 small-signal sensor wires (<100 mA each), in addition to these there are two silicone-tube gas-lines to provide propellant to the EPT. The total parasitic force load caused by these connections is minimized by taking them through the pendulum's center of rotation, as the flexure of each occurs with a small moment centered on the bearing axis. An unsecured slotted acetal axle spans the two opposing bearings and is press fitted into a bearing hub on each side (Fig. 6b). The press fit maintains axial rigidity but does not prevent the spreader-bar from correctly setting the separation distance between the two hubs (and hence, the bearing distance). The slot in the underside of the axel allows electrical and gas connections to exit the axel and drop to the pendulum plate at the desired locations. In this setup there is a small additional torsional parasitic force, determined by the path travelled by the wires and gas lines through the axel and where they terminate at either end.

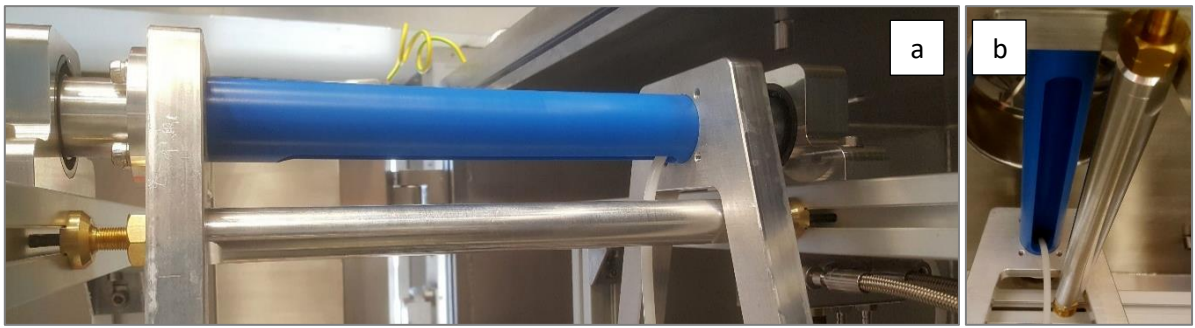

**Figure 6a**, Bearing, axel, and spreader bar design - used for bearing alignment, and **b**, wiring and gas feed slotted axel.

## 2.3 Instrument systems

At the time of writing eighteen custom instrument software modules are coordinated and controlled by an in-house supervisory control software system. A common software design methodology was employed to allow rapid development of new instruments and embed them in the supervisory control. This design framework incorporated features to allow easy integration of the following communication protocols RS-232, RS-485, USB, Ethernet, and GPIB. It was necessary to convert all non-ethernet communications to Ethernet near the instrument to reduce susceptibility to EPT noise. For both analogue and digital acquisition functions a National Instruments ethernet cDAQ chassis was used. Software control modules include:

- Weight calibration (RS-232)*
- Three-axis linear actuators (RS-232)*
- Pumping and vacuum chamber control (cDAQ, DIO)†
- Three-axis Hall sensors (cDAQ, AI)†
- Three-axis accelerometers (cDAQ, AI)†
- Load-cell (cDAQ, AI)†
- Philtech optical distancing (RS-232)†
- Pfeiffer turbo pump control (RS-232)
- Pfeiffer vacuum pressure readings (ethernet)

- 2 × 18 kW APMTech power supplies (ethernet)
- 2 × Alicat gas mass-flow controllers (RS-232)
- Lakeshore 218 temperature readout (RS-232)
- Lakeshore 224 temperature readout (ethernet)
- CryoTel CT cryocooler (RS-232)

*Further detail is given to these items as they are unique to these test facilities.

†Further detail is given to these items in the appendix for the benefit of the reader.

### 2.3.1 Weight calibration system

An integral part of most thrust-stands is the ability to do in-situ calibration of the sensor(s) used to make force measurements. For direct measurements this is most often undertaken using an adjustable weight arrangement that applies a known amount of force to the thrust-stand in a step-wise fashion [22,28,33,41]. A typical calibration method is to run the calibration cycle before a thruster test, and then again at the end, thereby allowing an assessment of the drift that may have occurred during thrust measurements. Drift is an insidious problem for most thrust-stands and has many causes, not all of which behave linearly with changes in temperature and pressure (two most common causes of drift). Some of the heat generated by the EPT will be propagated to the measurement apparatus and is often unavoidable. This heat can lead to thermal expansion of various mechanical parts, softening of plastics/rubbers, and can alter the stiffness of wire, in addition to numerous other effects which are detrimental to accuracy. Changes in pressure, i.e., vacuum conditions, prevent convective cooling, can cause shrinkage or distortion of porous components (including wiring systems) and prevent acoustic damping/dissipation of energy.

The weight calibration system shown in figure 7 incorporates a set of five hanging weights (interchangeable) that can be loaded and unloaded using a stepper motor driven cradle. The weights nominally hang on a stem connected to the thrust-stand via a 0.1 mm carbon fiber thread incorporating a full ceramic pulley arrangement. The far end of the thread is connected to fine-thread brass adjuster to ensure the correct hanging height. The cradle is a 3D-printed 316 stainless-steel part, the stem is made from a 3 mm diameter brass rod, the weights and conical stem locators are made from machined aluminum. The choice of materials used in the calibration system are expected to result in a minimal interaction with the strong magnetic fields generated by the HTS magnet. Three calibration weight sets have been made (five of each weight), and these are made up of individual weights of nominally, 1.39 g, 3.77 g, and 4.92 g (each weight is within ±0.005 g but has been individually weighed and numbered to within ±0.001 g). The resulting calibration ranges (using five weights) are 68 mN, 185 mN, and 241 mN. An automated software system is used to run calibration cycles, based on user defined: lifting heights, the lifting speed, and dwell times.

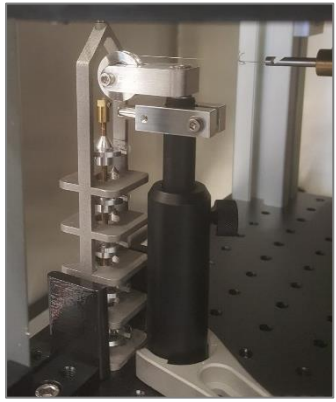

**Figure 7**, hanging calibration weights, pulley, and lifting cradle that sit beneath the pendulum plate.

### 2.3.2 Three-axis actuators

A three-axis, in-vacuum, actuator apparatus was designed and constructed (Fig. 8) to facilitate; i) in-plume measurements of thrust-force, ii) plasma diagnostics, and iii) to assess HTS magnetic field strength using a Hall sensor array. In the Z-axis (axial direction of the chamber) the maximum travel is 1 m, whereas the vertical (Y) and horizontal (X) axes have a maximum travel of 0.5 m, restricted by the circularity of the chamber. The actuators are controlled via three Pontech stepper motor drivers (operated in micro-stepping mode) allowing repeatable submillimeter movements. The software module used for controlling the actuators provides options for 1-, 2- and 3-dimensional manual and automated scans.

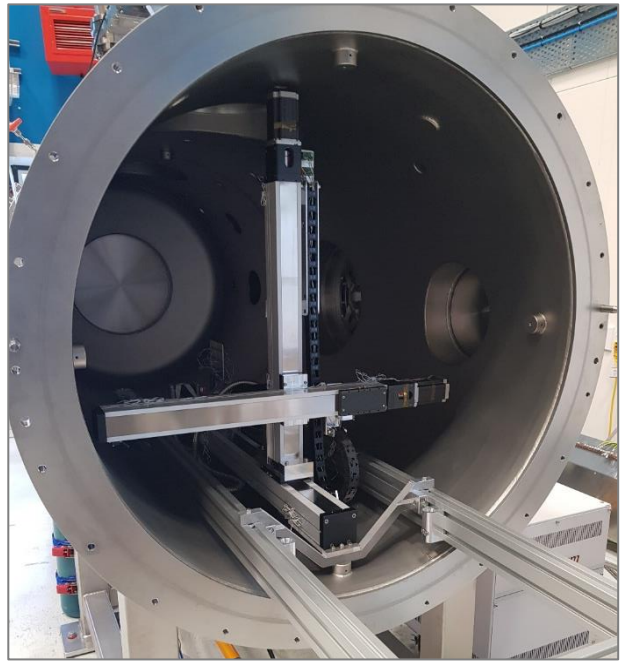

**Figure 8,** Three-Axis linear actuators mounted in chamber and canter-levered beam extending out of chamber for supporting the thrust stand.

## 3. Experimental methods

This section describes the initial thrust-stand testing methods that would be needed to establish the accuracy of a potential AF-MPD thruster measurement. The various components were tested in isolation, as part of building up a clear picture of how various subsystems influence the thrust-stand. The results of which are then used to build a tentative uncertainty budget. Note: at the time of writing the experimental data presented here only relates to in-air measurements, future work investigating in-vacuum effects.

### 3.1 Measurement capability

An important component test is the weight calibration system when operating on the unloaded pendulum, free from wiring, gas feeds and the water-cooling connections. These experiments ought to reveal the best measurement capability when making either a static (load-cell) or dynamic (optical displacement) measurement. Adjustment of the spreader-bar was used to align the pivot bearings, revealing the influence of bearing stiction on the scale linearity and hysteresis. Assessment of stiction is vital part of dynamic measurements where there are discrete pendulum displacements during both thrust measurements and weight calibration. Water cooling and wiring drops were then added to the thrust stand to evaluate the scale reduction due to parasitic force loads caused by these components.

### 3.2 Deadload performance testing

Another key performance test for the thrust-stand was deadload mass testing, where weights between 1 and 30 kg were added to the pendulum plate to quantify the reduction in scale resolution caused by a simulated increase in thruster weight. These dead-load tests were made using both the static and dynamic measurement systems. The AF-MPD under development is anticipated to weigh approximately 20 kg. However, assessing the thrust stands performance up to 30 kg provides valuable information on expected behavior for future thrusters that may be heavier.

### 3.3 Effects of cryocooler vibration

The cryocooler used to cool an HTS magnet introduces considerable vibration to the thrust-stand and thus, represents a potential loss in measurement capability. Therefore, thrust-stand performance testing includes quantitative measures of both the vibration and its impacts on force measurement accuracy. An aluminum and copper conductive cooling collar was added to the cryocooler hot end (Fig. 9a) to reject heat to the water chiller plate mounted on the underside of the pendulum plate. This collar had a stiffness sufficient to support most of the weight of the cryocooler in the horizontal orientation but was augmented with soft silicone isolators to support the rear section of the cooler and help dampen vibrations. Four PRTs were used to monitor temperatures, which included the cryocooler hot collar, cold-tip, pendulum plate, and the water chiller plate, as shown in figure 9b.

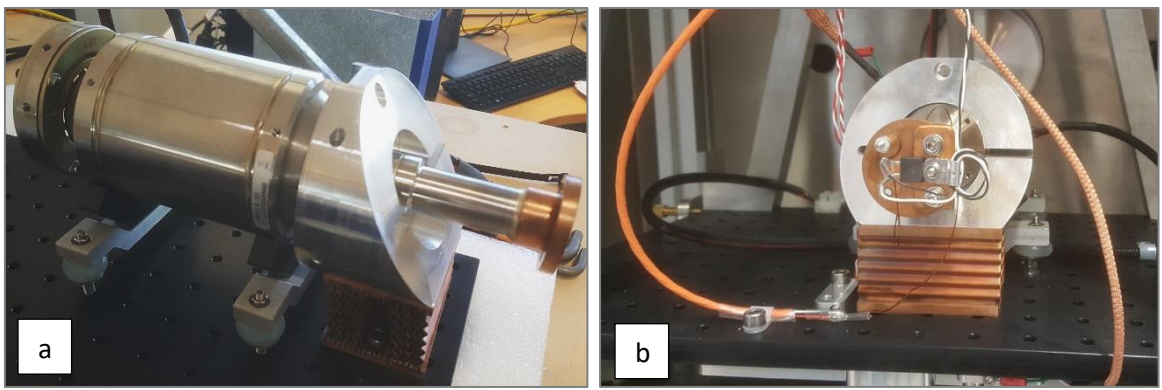

**Figure 9a**, CT cryocooler mounting and cooling components, and **b** Cryocooler mounted on pendulum plate with several PRTs used to monitor important temperatures at different locations.

# 4. Results and discussion

## 4.1 Verification of in-situ weight calibration system

Initial thrust stand experimental work was broken down into the following components tests:

- 4.1.1 Calibration sensitivity changes resulting from pivot-bearing misalignment
- 4.1.2 Calibration of the unencumbered pendulum
- 4.1.3 Calibration with the addition of water-cooling
- 4.1.4 Calibration with changing deadload
- 4.1.5 Calibration with cryocooler vibration

Components 4.1.1 – 4.1.4 were investigated in-air, whereas 4.1.5 was investigated in vacuum (due to cryocooler cold-head requirements). For calibration, components 4.1.1 – 4.1.3 used both the 68 mN and 185 mN scales, whereas in parts 4.1.4 and 4.1.5 only the 68 mN scale was used. Both optical and loadcell measurement techniques were used throughout these component tests.

### 4.1.1 Calibration sensitivity changes resulting from the pivot-bearing misalignment

In the first iteration of the thrust-stand only the load-cell and load-beam arrangement were fitted. Despite care in alignment of the two main thrust-stand pivot bearings, it quickly became evident that bearing stiction was varying with movement or loading of the thrust-stand. When the thrust-stand was statically loaded (using the load-cell) these small variations in stiction were hard to quantify and minimizing stiction was a tedious process undertaken by many repeated weight calibration cycles. However, with the addition of the optical displacement system the issue of stiction became easier to resolve, as any force applied to the thrust-stand would cause a displacement and thus, incite mechanical ringing in the under-damped pendulum. The magnitude and decay were seen to be directly related to the amount of pivot bearing striction. This observation led to the addition of the custom spreader-bar described in Section 2.2.2.

Given that the thrust-stand is under-damped, the process of spreader-bar adjustment requires the user to maximize mechanical ringing, which signals the point of the minimal stiction. Figures 10a and 10b show the improvements in optical displacement measurements that result from optimizing the bearing alignment (minimizing stiction). Bearing stiction was an anticipated problem and is one of the main reasons flexures and flexure-bearings are often preferred for many thrust-stand designs [31,33]. However, having the pendulum plate be i) intrinsically centered by gravity, ii) able to take high weight loads, iii) electrically isolated, and iv) impervious to magnetic fields were all determining factors in deciding to use a traditional ceramic bearing system. An additional benefit was being able to run cable and gas feeds through the bearing centers (reducing parasitic loading and not practical with flexure bearings). A valuable outcome from this method has been the demonstration that a traditional bearing pivot is a viable option for a simple hanging pendulum thrust-stand. Also, that tuning is possible enabling thrust measurements resolution of <1 mN.

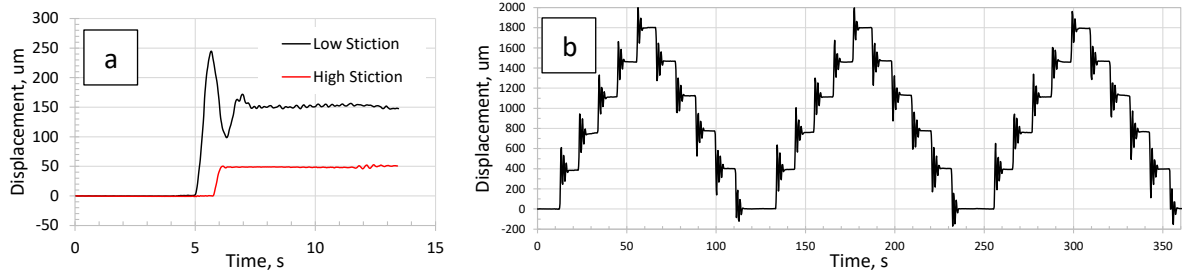

**Figure 10a**, Effect of spreader bar when used to minimize stiction and thus, maximize ringing (loading with single 1.39 g calibration weight, equivalent to 13.6 mN), and **b** Three successive calibration runs on unencumbered pendulum plate after spreader bar optimization (loading and unloading with five 3.77 g calibration weights, each equivalent to 37.0 mN).

### 4.1.2 Calibration of the unencumbered pendulum

Several calibration runs were made on the unencumbered pendulum system in air, as part of establishing base-line measurement capability. This entailed only having the weight calibration system attached to the pendulum plate and using either the load-cell (and load-beam) or optical measurement systems. Other than a small counterweight (<200 g), no other items were mounted on the pendulum. Calibrations were made using the 68 mN and 185 mN scales with three repeats of each. Figures 11a and 11b show respectively the typical calibration steps for the optical and load-cell measurements when using the 185 mN scale. Both plots also show the low pass filtering applied from which the calibration plateaus were derived. Visible in the optical measurements is the ringing associated with loading and unloading of each of the 3.77 g weights (not seen in the load-cell plot as it is statically loaded).

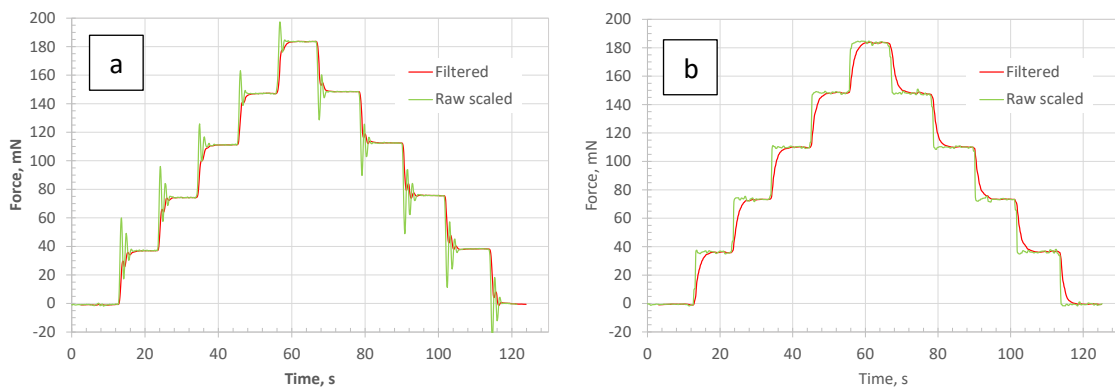

**Figures 11a and b,** Example of 185 mN Calibration data using optical displacement and via load-cell.

Table 2 summarizes data from these base-line set of experiments, derived from average plateau values over three runs, which were combined to determine maximum deviation from linearity (scale error), the standard deviation within a plateau for each weight, the maximum hysteresis (loading vs. unloading), and the drift from the start to the end of calibration. These values show the best measurement capability is approximately ±1 mN, without further data processing and analysis. It also reveals the increased hysteresis on the 185 mN scale due to stiction when using the optical displacement method. There was good repeatability for both measurement techniques when comparing individual plateaus over the three runs, typically with variations <0.5 mN. Drift values from the time of calibration start to end were also small and were <0.4 mN.

**Table 2**, Results from three successive runs using both the optical displacement (Opt.) and load-cell (LC) measurement systems on 68 mN and 185 mN scales.

| Experiment | Max. scale Error | Std. Dev. across three runs with varying number of weights | | | | | Max. Hysteresis | Drift |
|---|---|---|---|---|---|---|---|---|
| | | 1 | 2 | 3 | 4 | 5 | | |
| Opt. 68 mN | −1.16 | 0.42 | 0.24 | 0.17 | 0.35 | 0.16 | +0.60 | −0.02 |
| Opt. 185 mN | -1.05 | 0.46 | 0.15 | 0.20 | 0.33 | 0.32 | +1.41 | +0.17 |
| LC 68 mN | +0.87 | 0.45 | 0.05 | 0.21 | 0.34 | 0.25 | +0.66 | +0.37 |
| LC 185 mN | +0.77 | 0.41 | 0.35 | 0.35 | 0.33 | 0.05 | +0.55 | -0.09 |

The parasitic forces for the unencumbered thrust stand will be those contributed by the pivot bearings and by the pulley systems. For optical measurements a single pulley is used as part of the weight calibration system. However, when using the load-cell there is a second pulley that connects the pendulum plate and the load-beam. Therefore, in terms of quantifying parasitic force loads the optical measurement is more useful, as it only incorporates a single pulley. It could be argued that the load-cell arrangement limits movements of the pivot bearing, which ought to result in low stiction values, but quantifying this value may prove difficult without first evaluating the dynamic case.

A first order calculation based on the variables discussed in section 2.2.1 was used to assess the bearing and pulley stiction components, with the results given in Table 3. These specific calculations relate to the parasitic loading when making optical displacement measurements. The table provides the expected displacement for a frictionless pendulum subjected to a force acting at a radius arm equal to the pulley location (see Fig. 5). These show that the calculated and measured values differ by roughly 27%, which ought to be indicative of the combined effect of bearing stiction and pulley capstan friction. Part of future investigation will be to alter the small pulley configuration using different contact angles and an additional pulley to better quantify the pulley capstan component. From this work the pivot bearing stiction could be more accurately known and methods developed to minimize or achieve repeatable stiction through more precise spreader bar adjustment. However, this may be difficult, as we expect the pivot bearing stiction losses to be at least an order of magnitude larger than those of the capstan pulley, based on the pivot bearings larger size, weight loading, and contact area.

**Table 3**, Calculated optical displacement due to calibration weight.

| Parameter | Value |
|---|---|
| Radius, pendulum center-of-mass, m | 0.391 |
| Radius optical sensor, m | 0.592 |
| Radius of pivot bearing | 0.02 |
| Mass of pendulum and swing arms, kg | 8.706 |
| Gravity, m·s$^{-2}$ | 9.81 |
| Calibration Force, mN | 185 |
| Measured tilt angle, Radians | $2.24 \cdot 10^{-3}$ |
| Measured optical displacement, μm | 1325 |
| Calculated displacement at optical radius, μm | 1818 |
| Calculated frictional coefficient | $16.23 \cdot 10^{-3}$ |
| Torque due to calibration force, mN·m | 102.5 |
| Torque due to friction force, mN·m | 27.7 |
| Torque due to gravity, mN·m | 74.8 |
| Losses due to bearing stiction and pulley losses, % | 27.0 |

### 4.1.3 Calibration with the addition of water-cooling

Building on the unencumbered configuration, a water-cooling plate (Appendix 6.2) was then added to the underside of the pendulum plate and a flow rate of 1 L·min$^{-1}$ applied to it. A triple set of measurements were again made with both the optical displacement and load-cell systems over both calibration ranges, 68 mN and 185 mN, calibrations were also made with zero flow on the 68 mN scale. These measurements ought to reveal any performance changes due only to water connections and the additional weight from the cooling plate.

To investigate the static force load contributed by residual off-axis water flow a step change in the flow was made, going from 1 L·min$^{-1}$ to 0 L·min$^{-1}$. In this instance it was important to change between the 'flow' and 'no-flow' regimes. This order would limit any center of mass changes caused by thermal expansion or contraction of pendulum plate if the water and pendulum plate were at different temperatures. Measurements showed that switching between 1 L·min$^{-1}$ and 0 L·min$^{-1}$ resulted in a stepwise static force load change of +22 mN (optical) and +11 mN (load-cell). When this flow order was reversed, stepwise changes of −10 mN Optical) and −8 mN (load-cell) were seen with long settling times, not reaching equilibrium until several minutes later. This step change in force is caused by small differences in water-in and water-out flow force vectors, as the water passes through 12 mm diameter thin-wall soft-silicone tubes (Appendix 6.2, Figs. 18a and 18b) that traverse the gap between thrust-stand and pendulum plate. A lesser secondary effect will be caused by any small path length differences within the symmetrical chiller plate labyrinth.

Similar to the preceding section, a summarized table of triplicate results is shown in Table 4. It can be seen that the repeatability of loading weights and drift for both the optical and load-cell systems are similar to those seen for the unencumbered thrust-stand. However, values for the optical measurements reveal an increase in both the scale-error and hysteresis. For the optical 185 mN calibration, the Max. scale-error results were skewed by an outlier of the fourth calibration weight caused by an incorrect *set-height* value in calibration software. This height error resulted in a fractional pickup of the fifth calibration weight leading to a consistently high value during loading and unloading. Subsequent measurements suggest the Max. scale-error ought to be 2–3 mN. Despite this outlier, overall, these results show the optical displacement measurements are more affected by the chiller system when compared to the static load-cell arrangement.

**Table 4**, Results for water chiller plate at a flow rate of 1 and 0 L·min$^{-1}$, three successive runs using both the optical displacement (Opt.) and load-cell (LC) measurement systems on 68 mN and 185 mN scales.

| Experiment | Max. scale-error | Std. Dev. across three runs with varying number of weights | | | | | Max. Hysteresis | Drift |
|---|---|---|---|---|---|---|---|---|
| | | 1 | 2 | 3 | 4 | 5 | | |
| †Opt. 68 mN | +2.53 | 0.35 | 0.42 | 0.58 | 0.67 | 0.05 | +3.46 | −0.05 |
| †Opt. 185 mN | +3.99 | 0.12 | 0.19 | 0.15 | 0.17 | 0.40 | +1.87 | +0.28 |
| †LC 68 mN | +1.62 | 0.62 | 0.50 | 0.53 | 0.25 | 0.22 | +0.94 | +0.41 |
| †LC 185 mN | −1.30 | 0.43 | 0.38 | 0.33 | 0.35 | 0.13 | +1.37 | -0.01 |
| *Opt. 68 mN | +1.68 | 0.27 | 0.33 | 0.54 | 0.68 | 0.30 | +3.40 | +0.15 |
| *LC 68 mN | +0.74 | 0.13 | 0.40 | 0.70 | 0.64 | 0.43 | +0.38 | +0.01 |

† Measurements made at 1 L·min$^{-1}$
* Measurements made at 0 L·min$^{-1}$

## 4.2 Calibration with changing deadload

An important parameter to characterize is the impact of a thruster weight on the sensitivity of the thrust-stand. For most pendulum style thrust-stands there is a decreasing measurement sensitivity with increasing thruster weight. It is anticipated the AF-MPD thruster could weigh up to 20 kg. To build confidence in whether there is sufficient measurement resolution under this loading, calibrations were done with both the optical and load-cell systems to a maximum deadload of 30 kg using the smaller 68 mN calibration scale. The deadload weights of 1, 2, 5, 10, 15 and 30 kg were used, the order of which was then reversed to reveal any hysteretic behavior in the thrust-stand during mechanical relaxation. figure 13a shows the results for the optical measurements in units of microns, whereas figure 13b shows this same data but as a percentage loss in scale, together with the reversal of the deadload weights. A 70% loss of scale results from 30 kg and is about 65% at around 20 kg. These measurements also show the optical setup has low hysteresis with changing deadload. Despite a significant drop in resolution (going from 265 µm to 78 µm with a 30 kg deadload), the scale-error and hysteresis values were proportionally near identical to those seen for 0 kg, being respectively +2 mN and +2.5 mN after scaling was applied. This was consistent across the various deadload weights used, meaning the loss in resolution had little impact on the potential measurement accuracy. Therefore, hysteresis caused by dynamic friction in the pivot bearing can be treated as a near linear function of deadload weight.

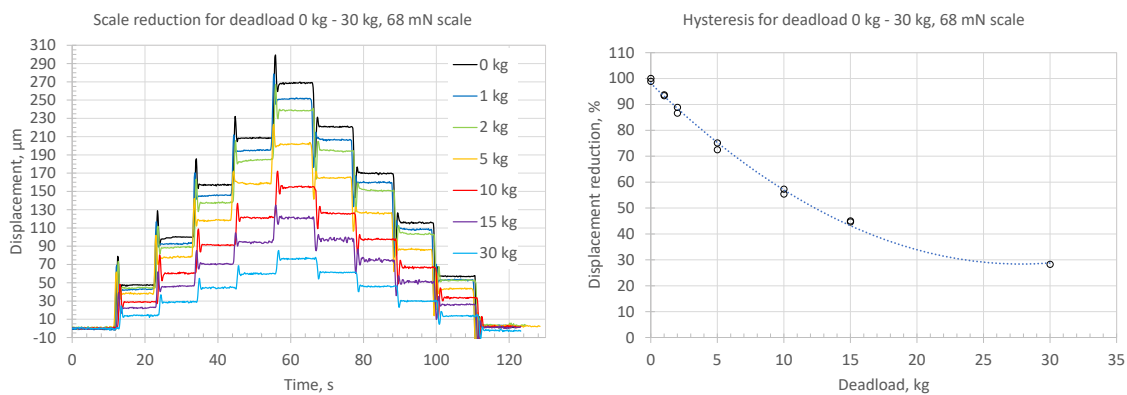

**Figures 12a, and b,** Deadload weight calibration data up to 30 kg (left) and percentage loss in scale (right) using the optical displacement configuration of the thrust stand.

Using the same methods described above data was also gathered for the load-beam and load-cell, so similar comparative loss in scale could be made. The results from both techniques could then be used to inform the most suitable thrust measurement method. Figures 13a and 13b show respectively the calibration data in millinewtons and percentage loss in scale. For context the load-cell voltage scaling and offset used for the 0 kg calibration run was also used for all subsequent deadload weight results. Two distinct observations can be made when comparing with the optical measurements of Figure 12. First, the scale reduction is smaller, only 60% at 30 kg (compared to 70% for the optical) and second, the deadload hysteresis is larger, approximately 4% (compared to <1% for optical). The increased deadload hysteresis is likely due to the need to use an additional in-line spring with the load-beam to prevent sudden or accidental loads from damaging the load-cell. Whilst adding and removing deadload weights this spring may have undergone small non-elastic extensions leading to the observed hysteresis. Unlike the optical configuration the scale error remained constant at around +1 mN, but calibration hysteresis values reduced with increasing deadload weight (after scaling), going from a maximum of about +2.5 mN (between 0 and 5 kg) and then steadily dropping to +0.5 mN at 30 kg. This behavior was repeated as deadload weights were removed from the thrust-stand. The change in hysteresis might be explained in terms of the pivot bearing friction. The load-cell arrangement is near static, whereas the optical system, being a displacement measurement, undergoes both static and dynamic friction.

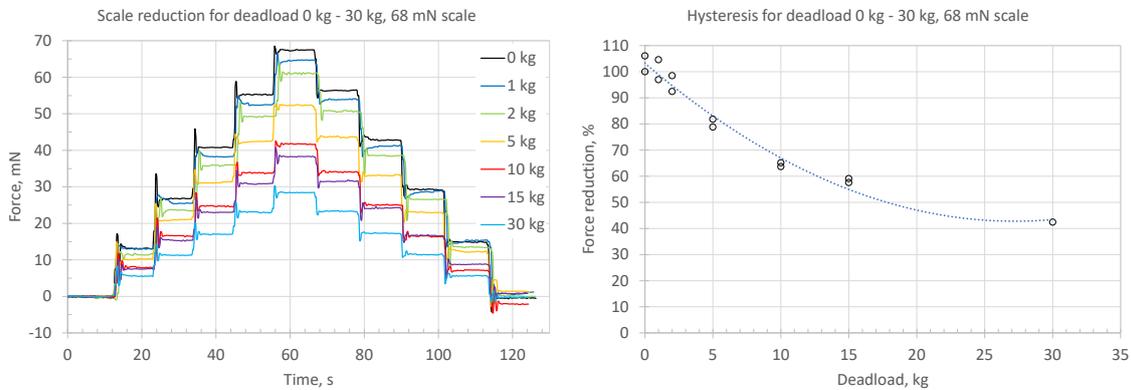

**Figures 13a,** Deadload weight calibration data up to 30 kg and **b**, percentage loss in scale using the load-cell configuration of the thrust stand.

For both the optical and load-cell measurements there is a near parabolic loss in sensitivity due to deadload weight. We expect the real relationship would be more accurately described by an equation incorporating $m \cdot g \cdot \sin\theta$, center-of-mass radius, and pivot friction terms (similar to calculations presented in table 3). However, that level of detail is omitted here for the sake of space.

### 4.3 Calibration with cryocooler vibration

In-vacuum vibration and calibration measurements were made while operating a SunPower CryoTel CT cryocooler configured horizontally, off-axis on the thrust stand (vibration in line with pivot axis). These experiments were performed using both the load-cell and optical systems on the 68 mN calibration scale with the water chiller flow set to 1 L·min$^{-1}$. Example measurements using both the optical and load-cell systems at 60 W and 140 W are shown in Figures 15a to 15d. These reveal the load-cell measurements are far more susceptible to drift when compared to the optical measurements. Also, that despite the apparent oscillatory behavior in the optical measurements, filtering allows good measures of the calibration plateaus. An interesting result from the optical series of experiments was the near elimination of loading-unloading hysteresis, which is thought to be caused by pivot bearing stiction. This is probably due to the settling effect of vibration from the cryocooler. This reduction in hysteresis was not seen in the load-cell measurements and may again be the result of a small damping spring used in-line with the load beam (as was done with the deadload experiments to protect the load-cell). Typical scale-errors and hysteresis values were less than 0.5 mN for the optical measurements at both 60 W and 140 W. For the load-cell scale-errors and hysteresis values were approximately 2 mN and 4 mN, respectively, at 60 W and 140 W.

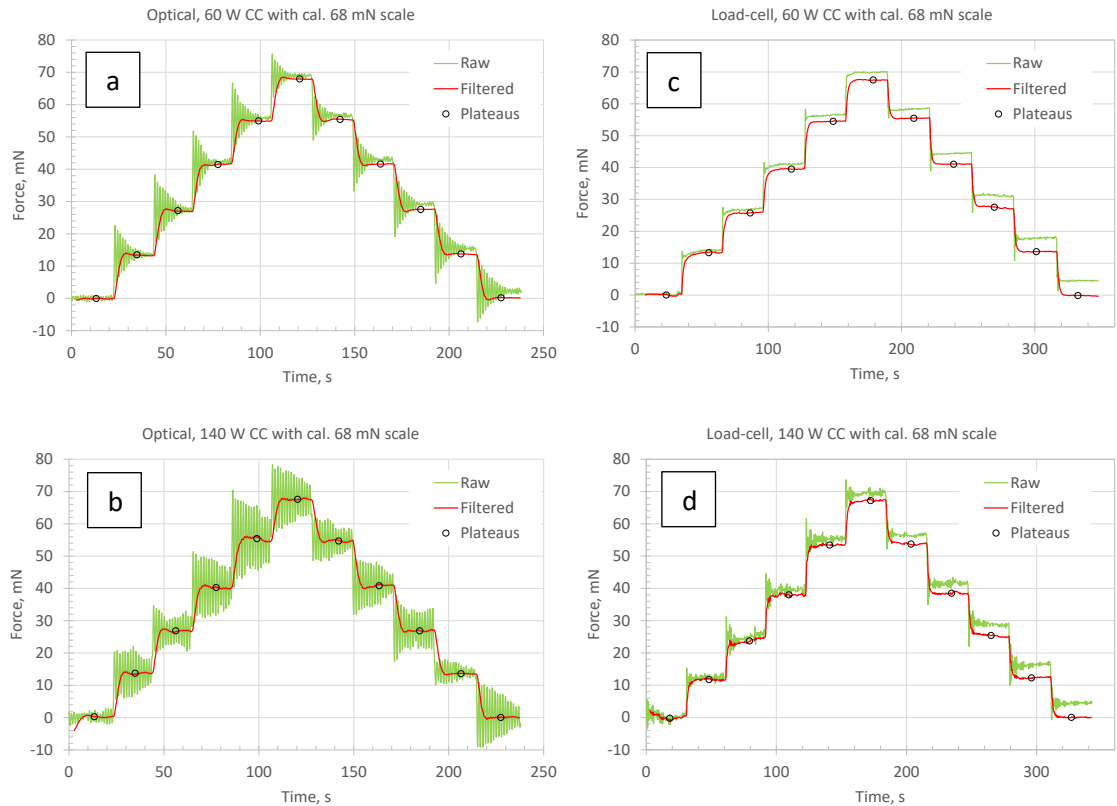

**Figures 14a, b, c, and d,** Example calibration plots for both optical (14a and 14b) and load-cell (14c and 14d) systems when subjected to a cryocooler operating at either 60 W (14a and 14c) or 140 W (14b and 14d).

Measurements of vibration were made using a three-axis Freescale FXLN8361Q accelerometer, and each axis was scaled against earth's gravity using a rotary vernier stage. Figure 15 reveals the sensitivity of the accelerometer setup when the weight calibration system lowers a 1.39 g (13.6 mN) weight onto the thrust-stand (cryocooler not running). Visible are the individual pulses from the stepper-motor as it lowers the cradle holding the weight. Directions can be interpreted as, x-axis in line thrust direction (horizontal), y-axis is vertical, and z-axis horizontal off-axis (in line with pivot).

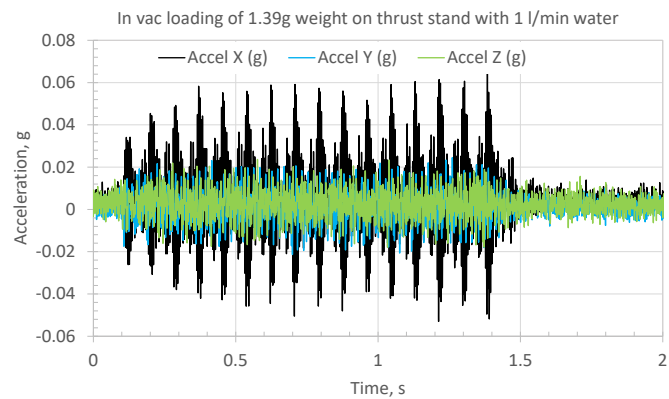

**Figure 15**, Example of vibration due to 1.39 g weight being loaded onto thrust-stand.

In stark contrast to the loading of a calibration weight is the cryocooler startup acceleration (40 W), shown in Figure 16, which shows the peak-to-peak acceleration is close to 6 g. This settles to around ±1 g after 1 second and is indicative of the acceleration seen by the thrust-stand while operating the cryocooler.

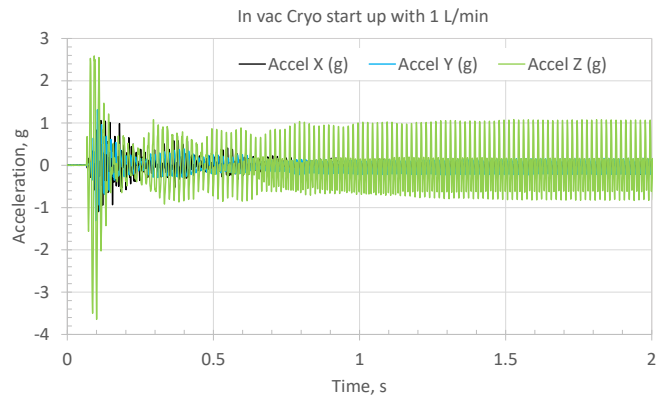

**Figure 16**, Acceleration seen on thrust-stand at cryocooler startup.

The cryocooler operates at 60 Hz and transmits some of its vibrational energy (not absorbed by an inbuilt passive damper) to associated mounts and surrounding equipment. A Fast-Fourier-Transform of the accelerometer data was performed to derive Power Spectral Density (PSD) information and is presented here [21]. Numerous higher order harmonics of similar magnitude to the fundamental (60 Hz) were seen in the x-axis (thrust direction), despite the care taken to use soft silicone vibration dampers on the base of the thrust-stand and cryocooler mounts.

### 4.4 Thrust-stand uncertainty budget

A basic measurement model can be used to describe the thrust-stand in terms of an individual measurement $w(i)$, the measurand ($Y$), and the various contribution uncertainties ($E_i...E_n$). A measurement equation can then be constructed in a summative form in regard to the measurand and uncertainty terms,

$$w(i) = Y + E_{rep} + E_{Hys} + E_{Fric} + E_{res} + E_{Drift} + E_{Scale} + E_{Rand} \qquad (1)$$

The following uncertainty components ($E_i...E_n$) have been considered,

- ($E_{rep}$) **Repeatability** is the ability to achieve the same plateau value for a given calibration weight from one calibration run to the next.
- ($E_{Hys}$) **Hysteresis** could be interpreted in two ways. The first, is comparisons made within a single calibration run, i.e., the difference in values measured between loading and unloading of the same weight. The second, involves comparing the same weight across multiple runs during either loading or unloading. For the purposes of this analysis, the first scenario is used, as it considers only a single calibration run and where the hysteretic uncertainty is attributed to stiction.
- ($E_{Fric}$) **Static friction** for a given a pendulum displacement (and thruster weight) the pivot bearings and pulley system will contribute a near constant frictional offset, with small variations related properties of those parts. However, in addition to this, the numerous mechanical and electrical connections to the thrust-stand will also impart a variable frictional load, that will be a function of both temperature and pressure.
- ($E_{res}$) **Resolution** of the optical system greatly exceeds the noise floor (±0.001 µm) and is mentioned here more for completeness rather than as a real uncertainty contribution.

- ($E_{\text{Drift}}$) **Drift,** is most often caused by changes to the thrust-stand's center of mass (COM), and results from a combination of thermal expansion, physical connections, and mechanical settling (e.g., from vibration isolation components). Thus far, these effects have behaved linearly over time. Therefore, it can be corrected for, during calibration and over the time taken to make a thruster measurement. At the time of writing only drift due to water cooling plate (thermal), cryocooler (thermal) and deadload (mechanical) changes have been observed. However, it is anticipated that this uncertainty component will need to be enlarged for AF-MPD measurements, where there will be several more items likely to affect COM, and hence drift.
- ($E_{\text{scale}}$) **Scale error** was present to some extent in all optical measurements, irrespective of thrust-stand loading or configuration. This error manifests as a small departure from the expected value with the cumulative loading or unloading of calibration weights, with the error being maximum at the third weight (mid-scale). A linear scaling factor based on zero and five weights is used to transform data from microns to millinewtons. A departure at the mid-point is therefore not unexpected and may be due to slight non-linearity in the optical decoding of signal to distance. Note, this error is not due to a small angle approximation error, as this has been calculated to be less than 0.001 microns (based on unencumbered pendulum measurements, where deflections are largest).
- ($E_{\text{Rand}}$) **Noise** sources in the optical measurements originate from vibration and other optical sources (external + optical sensor system), with vibration being the largest. The noise can be mitigated to some extent through longer averaging times, but this tends to then limit the system's response time to greater than 1 sec (approximately the response time of the pendulum).

For illustrative purposes a combined uncertainty is calculated for a non-thruster, non-magnet measurement, where we consider the set of measurements made with the cryocooler operating in-vacuum at 140 W (total pivot bearing load ~13 kg), the water-cooling plate with a flow rate of 1 L·min$^{-1}$, and with four four-wire PRTs each using a 4 mm diameter PTFE coated shielded cable. The measured displacement at the maximum calibration weight of 68 mN was 244 μm.

The various uncertainty components for this setup are shown in table 5. No allowance has been made for uncertainty in the weight calibration system, as the uncertainty budget is expected to be dominated by other mechanical parts. Uncertainty calculations were made via the usual method (simple) and via the more thorough Welch-Satterthwaite method:

*Simple method*  $\qquad U_{\text{total}}^2 = U_1^2 + \ldots U_N^2,$

where $U$ values are the expanded uncertainty components and their respective degrees of freedom when accessed at the 95% confidence level.

*Welch-Satterthwaite method*  $\qquad V_{\text{total}} = [\sqrt{(s_1^2 + \ldots s_N^2)}]^4 / (s_1^4/V_1 + \ldots s_N^4/V_N),$

Where $V_{\text{total}}$ is the total degrees of freedom, $s_1$ to $s_N$ are the standard uncertainty components and $V_1$ to $V_N$ are the individual degrees of freedom.

**Table 5**, Tentative uncertainty budget for thrust-stand with cryocooler running at 140 W in vacuum with water cooling.

| Uncertainty Component | Distribution Type | Degrees of freedom | Standard deviation, μm | Coverage factor, $k$ | Expanded Uncertainty, μm |
|---|---|---|---|---|---|
| Repeatability ($E_{\text{rep}}$) | Gaussian | 8 | 2 | 2.31 | 4.62 |
| Hysteresis ($E_{Stic}$) | Gaussian | 8 | 2 | 2.31 | 4.62 |
| Offset ($E_{Fric}$) | Gaussian | 24 | 3 | 2.06 | 6 |
| Resolution ($E_{res}$) | Gaussian | 8 | 1 | 2.31 | 1.2 |
| Drift ($E_{Drift}$) | Gaussian | 8 | 5 | 2.31 | 11.6 |
| Scale error ($E_{scale}$) | Gaussian | 8 | 5 | 2.31 | 11.6 |
| Noise ($E_{Rand}$) | Gaussian | 24 | 1 | 2.06 | 2 |
| **Total uncertainty for displacement measurements** | | | | | **Uncertainty, μm** |
| Total Expanded Uncertainty, simple | | | | | 18.8 |
| Total degrees of Freedom, $V_{total}$ | | | | | 29.1 |
| Total Expanded Uncertainty, W-S | | | | | 17.0 |
| Total Standard Uncertainty, W-S | | | | | 8.3 |
| **Total uncertainty when converted to force measurements** | | | | | **Uncertainty, mN** |
| Total Expanded Uncertainty (68 mN calibration scale with 244 μm of displacement, 95% confidence level, ±2σ) – Using simple uncertainty method. | | | | | ±5.2 |
| Total Expanded Uncertainty (68 mN calibration scale with 244 μm of displacement, 95% confidence level, ±2σ) – Using W-S method. | | | | | ±4.7 |
| Total Standard Uncertainty (68 mN calibration scale with 244 μm of displacement, 68% confidence level, ±1σ) – Using W-S method. | | | | | **±2.3** |

From Table 5 the expected standard uncertainty for the thrust-stand is about ±2.3 mN, which is surprising small when considering the 3.1 kg cryocooler generating ±1 g of 60 Hz noise and the water-cooling system. This uncertainty budget will be a need to be reevaluated as the thrust-stand is progressed for use an AF-MPD thruster with a superconducting magnet. However, it seems reasonable to believe based on current measurements that the uncertainty value can be kept well below ±5 mN when measuring thrust in the range of 50 – 100 mN.

## 5. Conclusions and future work

This paper has described a snapshot of progress towards an operational EPT test facility, purpose built for investigation of high-temperature superconducting (HTS) AF-MPD thrusters. A large part of this work is now complete with the results showing successful vacuum facilities and late-stage thrust-stand testing. It is anticipated the thrust-stand will continue to evolve with the integration between thruster and thrust-stand. These evolutionary changes will be presented in future works related to our first generation of HTS AF-MPD thruster, currently in its final stages of development.

*Vacuum systems*

An automated vacuum system has been demonstrated to provide rapid pump-down, venting and cryopump regeneration processes. The typical pump down time to reach $1\cdot10^{-5}$ hPa <2 h using a combination of a roots-blower scroll pump, turbo molecular pump and cryopump. Whilst operating argon flow rates in the range of 2 – 5 mg·s$^{-1}$ the background pressure can be kept to around $5\cdot10^{-4}$ hPa, which is sufficient for initial AF-MPD thruster testing.

*Thrust-stand*

A purpose built thrust-stand was constructed to be insensitive to the high-vibration and strong magnetic fields associated with HTS AF-MPD thrusters. Respectively, these effects are generated by a Cryo-Tel CT cryocooler (used to cool the magnet to ~60 K) and an HTS magnet capable of generating more than 1 T fields. The thrust-stand was shown to withstand cryocooler startup vibrations with a peak-to-peak amplitude of ~6 g and a continuous vibration of ±1 g. Under these conditions successful calibrations were made that had an uncertainty value of ±2.3 mN (68 mN calibration range). At the time of writing experiments to investigate the interactions between magnetic field and thrust-stand are still in progress, but so far, they show only a small and linear offset during HTS operation. Two different thrust measurement methods were investigated on the thrust-stand, one using an optical displacement system and the other using a load-beam and load cell arrangement. Despite the optical system having slightly larger hysteresis values during calibration cycles it was shown to be more robust and better suited to HTS AF-MPD thruster measurements.

*Facilities and Instrument systems*

An easy to use, one-person operable experimental test facility has been built. The main 3 $m^3$ vacuum chamber incorporates a rail system, which ensures a high level of safety (electrical and mechanical), ease of access, and rapid turnaround times when breaking chamber. A large suite of interconnected, but also standalone software modules were made to communicate with and control the various instrument systems. These modules are in turn coordinated by a supervisory control system. This control system can currently accommodate up to eighteen disparate instrument systems running a range of instrument interfaces, that include USB, Ethernet, GPIB, and Serial.

*Future activities*

There are several areas where it would be desirable to increase or improve the current test facilities capabilities, these include:

- **In-plume thrust measurements:** To allow measures of the impulse and transient thrusts. Traceability would be made via steady state thrust measurements comparisons between direct and indirect methods.
- **Plasma diagnostics:** Langmuir probes are being developed to facilitate plasma diagnostics using the three-axis actuators.
- **Thrust swirl measurements:** An interesting and potentially useful addition would be the ability to measure the thrust swirl component, inherent in all AF-MPD thrusters. These measurements would help determine the proportion of thrust lost to unwanted torque while also aiding in future thruster development work.
- **Thermal center of mass (COM) modelling:** Thermal COM changes during EPT operation are a problem for most thrust-stands. Techniques like finite element modelling could help in estimating these changes and reducing uncertainties.
- **Magnetic field mapping:** Field mapping has begun using the three-axis actuators for a small number of fields strengths, but more scans are needed to build a comprehensive data set that could be used for modelling thruster-magnet interactions in addition to understanding chamber-magnet effects.
- **Meisner Cold-traps:** Calculations recommend background pressures of $<1\cdot 10^{-5}$ hPa and preferably $<1\cdot 10^{-6}$ hPa during thruster operation. Custom Meisner cold-traps have been built, but not yet commissioned, with the intention of achieving such pressures in GERALDINE during operation of the AF-MPD.
- **Pressure distribution modelling:** Existing literature suggests there may be significant pressure distributions present near the EPT during testing, meaning the pressure gauge readings might not represent an accurate measure of the background pressure near the thruster. Modelling of gas flow, pumping locations and rates, in conjunction with chamber dimensions should give insights as to the magnitude of this problem.

- **Beam dump:** A graphite plate will soon be added down-stream of the thruster exhaust to mitigate sputtering of chamber material by the plume.

## Acknowledgments
This work was funded by the New Zealand Ministry of Business, Innovation and Employment through the project "Electric Propulsion for space", contract number RTVU2003.
.

# 6. Appendix

## 6.1 Vacuum chamber rail system

The main 3 m³ vacuum chamber interior is accessed by moving a 750 kg, 0.5 m long end-section on a set of custom-made rails, figure 17. The rails allow for 1.5 m of travel, giving routine access to a large portion of the chamber interior.

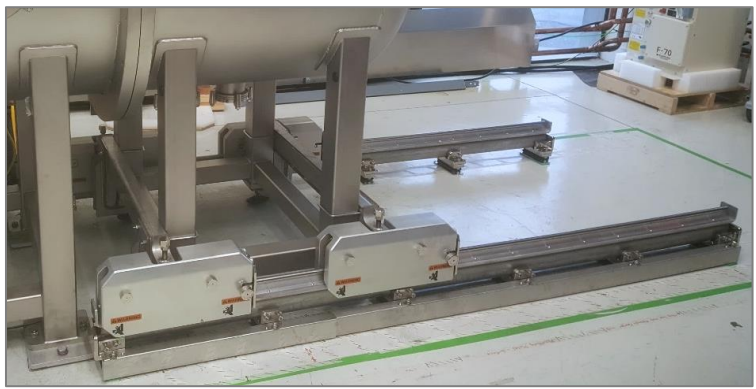

**Figure 17**, Rail system for chamber end section removal to allow the chamber to be one person operable.

## 6.2 Water cooling plate

A symmetrical labyrinth water cooling plate was added to the underside of the thrust-stand pendulum plate to aid in heat removal from several systems that are to be secured to it. The main sources of excess heat are from the cryocooler hot-end (180 W), optional Peltier cooler (hot-side, 50 W), and the anode of the HTS EPT (~300 W). The cooling plate has a maximum flow of ~2 L·min⁻¹, limited by the internal water-gallery dimensions, the ¼ inch Swagelok piping, and the head pressure of the laboratories main water chiller system (between 5 and 7 Bar). The water enters at about 20 °C and it is desirable to have it exit below 25 °C, to prevent undesirable temperature gradients on the pendulum plate. This results in a maximum cooling power of around 700 watts ($q = C_p \cdot m \cdot \Delta T$, where $q$ is heat in watts, $C_p$ is the heat capacity of water, 4186 J·kg⁻¹, $m$ is the mass flow of water, and $\Delta T$ the differential temperature between inlet and outlet, 5 °C).

The water enters and exists the chiller-plate via two large diameter soft silicone hoses (12 mm OD 10 mm ID), these are aligned with the axis of rotation as shown in figure 18a. This configuration combined with the symmetrical flow path within the cooling plate minimizes unwanted forces generated from the circulating water. Thus, allowing a range of flows to be used with very little impact to the thrust-stand performance. Several large 100 W stick-on heater pads were used to test the cooling-power of the chiller plate with an example configuration using two coolers is shown in figure 18b.

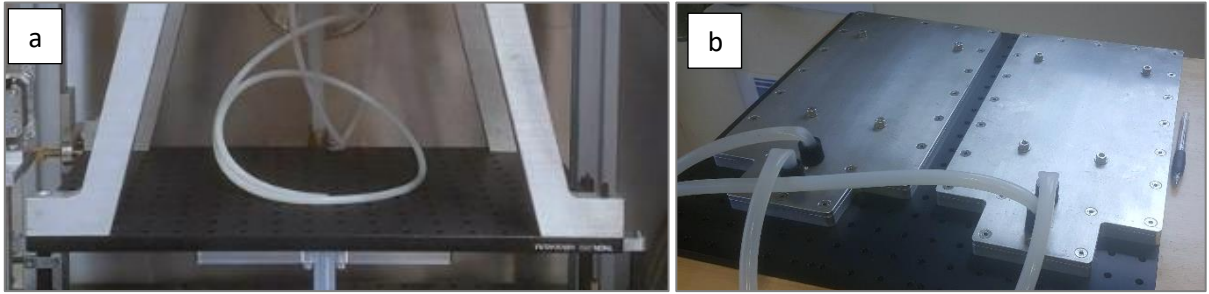

**Figure 18a**, Cooling plate mounted to underside of pendulum plate, and **b**, two identical water-cooling plates mounted to bread-board for testing cooling performance.

### 6.3 Dynamic measurement system

Dynamic displacement measurements are made using an optical Philtech DMS-D64 unit, which has a near field measurement range of 0 – 300 μm and a far field range of 500 – 4000 μm. This device employs an optical transceiver (mounted to the underside of the thrust-stand) to output a divergent infrared beam that is reflected off a nearby mirror (mounted on the thrust-stand pendulum). The returning beam will either under-fill (near-field) or over-fill (far-field) an array of optical fibers within the transceiver unit. By design, the near-field linear region has a sensitivity approximately ten times that of the far field, but the useful displacements range is correspondingly small. Close to the transition, from near- to far-field (300 – 500 μm) the output of the sensor is highly non-linear and zero at the transition, and as such this range must be avoided. Some force measurements have resulted in displacements >2000 μm. Therefore, the D-64 unit is normally setup to measure in the far-field. When also considering optical and mechanical noise this setup allows for displacement measurements with a resolution of about 1 μm. Given that the optical measurement radius is 554 mm, this results in an angular sensitivity of about $1 \cdot 10^{-4}$ degrees. The fiberoptic transceiver and mirror arrangement, when mounted on the thrust-stand can be seen in the lower region of figure 3b.

### 6.4 Static Loadcell system

To measure force statically a Honeywell FSS005WNSX load-cell rated to 5 N with a glass substrate was chosen. This load-cell design has a small deflection under maximum load, low-hysteresis, and high-linearity when compared to more tradition metal substrate-based units. However, the small size of the sensor (approx. 5 mm x 9 mm) necessitated it being mounted in a more practical mechanical configuration. A 3 mm thick 40 mm square precision machined aluminum front plate was used to locate the load-cell in addition to a miniature PRT for monitoring the load-cell temperature. This aluminum plate was backed with a 30 W Peltier unit to provide (optional) active temperature control of the load-cell during AF-MPD testing. The hot side of the Peltier unit was then backed with a more substantial 10 mm thick aluminum plate that allowed the front plate, load-cell, Peltier, and back plate to be sandwiched together using four M3 countersunk screws (Figures 19a and 19b). The back plate also provides a cooling path and mounting holes to allow the complete load-cell unit to be secured in the load-beam assembly shown in Figure 4. A PTFE insert is used to locate the Peltier unit, to prevent distortion of the top plate during assembly, and to provide thermal isolation between the front and rear of the Peltier unit.

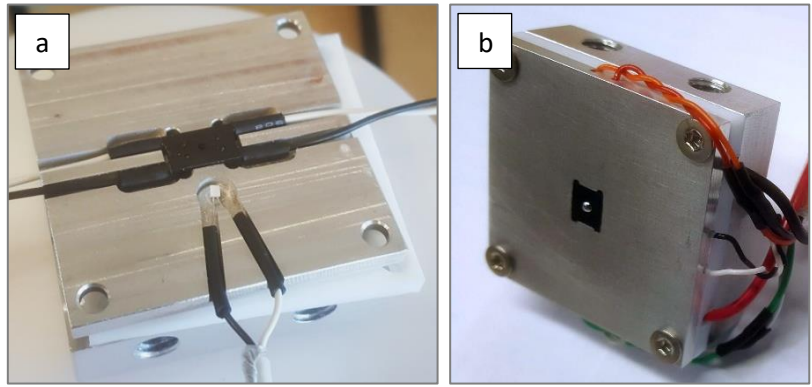

**Figures 19a**, Loadcell mounted in front face housing with embedded PRT), and **b,** when fully assembled with insulators and Peltier unit.

The load-beam, anvil, and pivot assembly have been designed to provide a twenty-fold mechanical amplification to the force applied by the load pulley on the underside of the pendulum plate. Hence, a force of 0.1 mN at the pulley (arbitrary measurement resolution) would result in a measured force at the loadcell of approximately 3 mN (ignoring small losses from the capstan idler pulleys and load-beam anvil friction). The load-cell has a sensitivity of 7 mV/V/N and when driven at 10 V this gives a sensitivity of 70 mV/N. Assuming an arbitrary resolution of 3 mN, the output from the load-cell would be around ~210 µV. The load-cell output is connected to a 24-bit NI-cDAQ 9219 analogue input device operating on the ±1 V range, giving a theoretical resolution better than 1 µV. Therefore, the load-cell and load-beam arrangement ought to have the necessary resolution required for sub-mN measurements.

Under static loading of the thrust stand any pendulum plate displacements will be considerably smaller than for the optical measurement system. Hence, the frictional losses and hysteresis values are also expected to be lower. For example, the loadcell data sheet suggests a 28 µm displacement under a 5 N load, which can be used to derive a displacement of ~34 µm at the pulley for a thrust force 200 mN. For comparison a similar optical measurement has a displacement of ~1500 µm (assuming an unloaded pendulum plate).

### 6.5 Automated vacuum control

As with many large vacuum facilities, automation of the pumping and venting operations is an important consideration, mostly for reducing risk to people and potential damage to equipment. Other benefits include rapid turnaround times, minimization of user intervention, and elimination of monitoring requirements. The use of a cryopump and cold-traps, which must not be exposed to atmospheric conditions, was also a contributing factor in the decision to automate our vacuum operations.

The control software module for the vacuum system is able to operate an ACG600 backing pump, a dry nitrogen supply for priming the back pump, a dry air supply for venting the chamber, a gate valve connecting the ACG600 pump to the chamber, a bellows valve to allow the ACG600 pump to operate on the cryo cavity (housing the Trillium CP16 cryo-head), a gate valve isolating the Trillium CP16 cryo-head from the chamber, and a gate valve used to isolate a ATH1603M turbo-pump from the chamber. A sperate software module is used to communicate with and control the ATH1603M turbo-pump. The Trillium CP16 cryopump must be manually switched on and off for equipment safety reasons.

Four distinct automated operations are possible and require minimal user input, these are: pump-down, chamber vent, cryopump cool-down, and cryopump warm-up. The automation sequences track the following items, cryo-head temperature, chamber pressure, backing pressure, and watch-dog timers – used to ensure the system remains within safe operating conditions. The turbo-pump can be engaged as a boost pump (by opening the corresponding gate valve) when the turbo-pumps temperature, speed, and the pressure inside the chamber are within safe limits.

### 6.6 Hall sensors

Three AHS P15a hall sensors were mounted orthogonally and electrically in series on a small G10 cube to allow simultaneous 3-axis magnetic field strength measurements. The sensors are power by a precision low-noise power supply (International Power, IHB28-1) coupled to an in-house Linear Devices LT3092 current source circuit designed to provide 1 mA. The sensors were calibrated using a Quantum-Design Physical Property Measurement System to generate fields between +/- 1 T and at temperatures between 288 and 303 K. The G10 mounting block is designed to be attached to the linear actuators described in Section 2.3.2 and their output read by a NI-9239 high speed 24-bit analogue input module. The software module for the hall sensors has a logging feature, which also records the actuator position to facilitate correlation of position and field strength.

### 6.7 DAQ systems

An NI-cDAQ 9189 network accessible chassis is used to provide digital IO for vacuum automation control, low-speed multi-function analogue inputs (NI-9219) used for load-cell and temperature measurements, high-speed analogue inputs (NI-9239) used for accelerometers and Hall sensor measurements. Each software module poles the NI-9189 chassis as required, with channels dynamically assigned based on currently installed DAQ modules and predefined configuration files associated with each software module.

### 6.8 High-power supplies for AF-MPD cathode and anode

Two 18 kW APMTech power supply units (PSU), each capable of 750 V DC and 65 A, are used to provide the power needed for ionization of the Ar propellant. The PSUs were chosen to provide a wide range of voltage and current options, thus, ensuring a range of potential AF-MPD thrusters could be designed and tested. For example, the HTS AF-MPD under development has a cathode requiring between 200 V and 500 V DC at a current of 0.6 A to 2 A and an anode requiring between 50 V and 300 V DC at currents of 2 to 15 A. Each PSU is remotely controlled by a standalone software module (via an ethernet link) and able to run automated voltage and current sequences.

### 6.9 Mass flow controllers

Two high-accuracy calibrated Alicat MC-500SCCM-D mass flow controllers are used to regulate the amount of Ar (propellant for AF-MPD). These controllers have a useable range between 0 and 12 mg·s$^{-1}$. Both manual and automated gas flows are possible using two separate software modules (one for the cathode and another for the anode). Automation can be achieved by software links to the high-power PSU outputs or by user defined timers.

### 7.0 HTS magnet cryocooler

A CryoTel CT sterling cryocooler is used to cool the HTS magnet on the AF-MPD and must be able to sustain a magnet temperature of <70 K, whilst rejecting ~180 W of heat from the hot end. The cryocooler represents a large vibration source, which will impact the thrust-stands measurement capability, coupled to its cooling requirements, and is discussed further Section 3.3. A software module is used for monitoring and control of the CT unit, providing feedback on the operating power and cold-head temperature.